\documentclass[manuscript]{aastex}

\def \ms{m\,s$^{-1}$\,}
\def \kms{km\,s$^{-1}$}

\slugcomment{draft: hd115617\_rev13.tex}

\shorttitle{Three low-mass planets orbiting 61~Virginis}
\shortauthors{Vogt et al.}

\begin{document}

\title{A Super-Earth and two Neptunes Orbiting the Nearby Sun-like star 61~Virginis}

\author{Steven S. Vogt\altaffilmark{1}, Robert A. Wittenmyer\altaffilmark{3}, R. Paul Butler\altaffilmark{2}, Simon O'Toole\altaffilmark{5}, Gregory W. Henry\altaffilmark{7}, Eugenio J. Rivera\altaffilmark{1}, Stefano Meschiari\altaffilmark{1}, Gregory Laughlin\altaffilmark{1}, C. G. Tinney\altaffilmark{3}, Hugh R. A. Jones\altaffilmark{5}, Jeremy Bailey\altaffilmark{3}, Brad D. Carter\altaffilmark{6}, Konstantin Batygin\altaffilmark{8}}

\altaffiltext{1}{UCO/Lick Observatory, University of California, Santa Cruz, CA 95064, USA}
\altaffiltext{2}{Department of Terrestrial Magnetism, Carnegie Institution of Washington, 5241 Broad Branch Road, NW, Washington, DC 20015-1305, USA}
\altaffiltext{3}{Department of Astrophysics, School of Physics, University of New South Wales, NSW 2052, Australia}
\altaffiltext{4}{Anglo-Australian Observatory, P.O. Box 296, Epping, NSW 1710, Australia}
\altaffiltext{5}{Centre for Astrophysical Research, University of Hertfordshire, HatÞeld, AL10 9AB, UK}
\altaffiltext{6}{Faculty of Sciences, University of Southern Queensland, Toowoomba, Queensland 4350, Australia}
\altaffiltext{7}{Center of Excellence in Information Systems, Tennessee State University, Nashville, TN 37209, USA}
\altaffiltext{8}{Department of Geological and Planetary Sciences, Caltech, Pasadena, CA 91125, USA}

\begin{abstract}

We present precision radial velocity (RV) data that reveal a multiple exoplanet
system orbiting the bright nearby G5V star 61~Virginis.  Our 4.6 years of
combined Keck/HIRES and Anglo-Australian Telescope precision RVs indicate the hitherto
unknown presence of at least three planets orbiting this well-studied star.
These planets are all on low-eccentricity orbits with periods of 4.2, 38.0,
and 124.0 days, and projected masses ($M\sin{i}$) of 5.1, 18.2, and 24.0
$M_{\oplus}$, respectively. Test integrations of systems consistent with the
RV data suggest that the configuration is dynamically stable.
Depending on the effectiveness of tidal dissipation within the inner planet, the
inner two planets may have evolved into an eccentricity fixed-point
configuration in which the apsidal lines of all three planets corotate. This
conjecture can be tested with additional observations. We present a 16-year
time series of photometric observations of 61~Virginis, which comprise 1194
individual measurements, and indicate that it has excellent
photometric stability.  No significant photometric variations at the periods of
the proposed planets have been detected. This new system is the first known
example of a G-type Sun-like star hosting a Super-Earth mass planet.
It joins HD~75732 (55~Cnc), HD~69830, GJ~581, HD~40307, and GJ~876
as a growing group of exoplanet systems that have multiple planets orbiting with
periods less than an Earth-year. The ubiquity of such systems portends that
space-based transit-search missions such as KEPLER and COROT will find
many multi-transiting systems.

\end{abstract}


\section{Introduction}

Over 400 extrasolar planets are now known.  The majority have been discovered by
using precision radial velocities (RVs) to detect the reflex barycentric motion of the
host star.  We have had a large sample of over 1000 nearby stars under precision
RV survey for the past 13 years at Keck and for the past 11 years
by the Anglo-Australian Planet Search (AAPS) at the Anglo-Australian Telescope
(AAT).  Of particular interest are the nearest brightest stars, as they allow
one to achieve the highest precision on stars worthy of follow-up with
space-based missions such as {\it Hubble Space Telescope (HST)}, {\it Spitzer}, and {\it James Webb Space Telescope (JWST)}.  Of equal importance to
stellar apparent brightness in RV precision is observing cadence.
Keplerian signatures are mostly strictly periodic, and the signal-to-noise ratio (S/N) of the detection
can be enhanced in the presence of a random noise background through increased
cadence.  However, obtaining adequate cadence on large telescopes like the Keck
and the AAT is quite difficult, as observing time is limited for any group.

Our groups have been combining observations from both the Keck and AAT on select
stars in the declination overlap region.  One of the target stars is
61~Virginis (61~Vir), a very nearby G5V star, only 8.5 parsecs away.  This star
was put on the Keck program in 2004 December and added to the AAPS target list
about four months later.  Over the past five years, we have accumulated a total of
206 precision RVs that indicate a system of at least three planets
orbiting this star. In this paper, we present all of these RV data
and discuss the planetary system that they imply.

\section{Basic properties of the host star 61~Vir}

61~Vir (=HD~115617, HR~5019, HIP~64924) is a very bright ($V$=4.74) and very
well-studied star.  It has a spectral type of G5V and lies at a distance of
only 8.52$\pm$0.05 pc \citep{per97}.  This star has been characterized by a
number of studies, with properties of interest to planetary system
characterization listed by \citet{vaf05} and \citet{sou08}.
Table~\ref{stellarparams} summarizes recent determinations of the fundamental
stellar parameters for 61~Vir.  Taken together, the properties of 61~Vir
indicate that it is an old, inactive star ideally suited for precision
RV planet searches.  Using Ca~H+K measurements taken between 1994
and 2006, \citet{hall07} found that 61~Vir is one of 13 targets for which the
observed variability is zero within the uncertainties.  In 29 observations over
7 seasons, they measure a mean $\log{R'_{\rm HK}}$ of $-4.93$.  \citet{henry96}
also measured activity in this star, and they found $\log{R'_{\rm HK}}=-4.96$,
which is consistent with Hall et al.\ above.  Similarly, \citet{Wit06} derived
a mean $\log{R'_{\rm HK}}=-5.03$ from five years ($N=18$) of observations at
McDonald Observatory.  Our measurement of $\log{R'_{\rm HK}} = -4.95$ leads to
an estimate \citep{jtw05} of 1.5 \ms\ for the expected RV jitter
due to stellar surface activity.  \citet{baliunas96} measured a rotation period
of $P=29$ days for 61~Vir.  The age of 61~Vir was estimated as
6.3$^{+3.3}_{-3.1}$ Gyr by \citet{vaf05} and 8.96$^{+2.76}_{-3.08}$ Gyr by
\citet{takeda07}.  In summary, 61~Vir is a nearby, bright, solar-type star
with physical properties quite similar to our own Sun.

\section{Observations}

The HIRES spectrometer \citep{vog94} of the Keck I telescope and the UCLES
spectrometer \citep{diego09} of the AAT have been used to monitor 61~Vir.  A
total of 126 AAT observations,  dating from 2005 April 21 to 2009 August 14,
representing a time span of 1576 days, have been obtained.  The median internal
velocity uncertainty for these AAT data is 0.65\,\ms. A total of 80 Keck
observations dating from 2004 December 29 to 2009 Aug 09, representing a data
span of 1685 days, have also been obtained.  The median internal velocity
uncertainty for these Keck data is 0.54\,\ms.

Doppler shifts were measured  \citep{but96} by placing an
Iodine absorption cell just ahead of the spectrometer slit in the converging
beam from the telescope.  This gaseous Iodine absorption cell superimposes a
rich forest of Iodine lines on the stellar spectrum, providing a wavelength
calibration and proxy for the point spread function (PSF) of the spectrometer.
The Iodine cell is sealed and temperature-controlled to 50 $\pm$ 0.1 $^{\circ}$C (at Keck)
and 60 $\pm$ 0.1 $^{\circ}$C (at the AAT) so that the column density of Iodine remains
constant.  For the Keck planet search program, we operate the HIRES spectrometer
at a spectral resolving power $R\approx70,000$ and wavelength range of
3700-8000\,\AA, though only the region 5000-6200\,\AA\ (with Iodine lines) is
used in the present Doppler analysis.  For the AAT program, we typically achieve
a spectral resolving power of $R\approx50,000$.  Doppler shifts from the
spectra are determined with the spectral synthesis technique described by
\citet{otoole08}.  The Iodine region is divided into $\sim$700 chunks of
2\,\AA\ each.  Each chunk produces an independent measure of the wavelength,
PSF, and Doppler shift.  The final measured velocity is the weighted mean of the
velocities of the individual chunks.

Table \ref{115617vels} lists the complete set of 206 RVs for
61~Vir, corrected to the solar system barycenter.  The table lists the JD of
observation center, RV, and internal uncertainty.  No offset has
been applied between the AAT and Keck in this table.  The internal
uncertainty reflects only one term in the overall error budget and results from
a host of systematic errors from characterizing and determining the PSF,
detector imperfections, optical aberrations, effects of under-sampling the
Iodine lines, etc.  Two additional major sources of error are photon statistics
and stellar jitter. The latter varies widely from star to star, and can be
mitigated to some degree by selecting magnetically inactive older stars and by
time-averaging over the star's unresolved dominant asteroseismological $p$-mode
oscillations. Since the single exposures required to reach a required S/N for
bright stars like 61~Vir are much shorter than the characteristic time scale of
low-degree surface $p$-modes, short exposures (ie., $<$5\,minutes) will add an
additional noise (or ``jitter'') component. This latter effect was recognized as
a potential noise source by the AAPS some time ago, and so, since 2005 July,
AAPS observations of bright targets like 61~Vir have been extended to the
10\,--\,15\,minutes periods required to average over these $p$-mode oscillations
\citep{otoole08,16417paper}.  For most of the past four years, only single
exposures of $\sim$7 s at Keck were taken of 61~Vir at each epoch, though in
2008 July we began $p$-mode averaging by combining multiple shots of 61~Vir over a
5\,--\,10 minutes dwell at each epoch.  All observations are then further binned on
two-hour timescales after being precision Doppler processed.

\section{Photometry}

In addition to our RV observations from Keck I and AAPS, we
have acquired high-precision photometric observations of 61~Vir during
17 consecutive observing seasons from 1993 April to 2009 April with the
T4 0.75~m automatic photometric telescope (APT) at Fairborn Observatory.
Our APTs can detect short-term, low-amplitude brightness variability in
solar-type stars due to rotational modulation of the visibility of surface
magnetic activity (spots and plages), as well as longer term variations
associated with the growth and decay of individual active regions and the
occurence of stellar magnetic cycles \citep{henry99}. The photometric
observations help to establish whether observed RV variations
are caused by stellar activity or planetary reflex motion
\citep[e.g.,][]{hbd+2000}. \citet{queloz01} and \citet{paulson04} have
presented several examples of periodic RV variations in
solar-type stars caused by photospheric spots and plages. The photometric
observations are also useful to search for transits of the planetary
companions \citep[e.g.,][]{hmbv2000}.

The T4 APT is equipped with a precision photometer based on an electromagnetic interference (EMI) 9924B
bi-alkali photomultiplier tube that measures photon count rates successively
through Str\"{o}mgren $b$ and $y$ filters.  The APT measures the difference
in brightness between a program star and nearby constant comparison stars.
Our automatic telescopes, photometers, observing procedures, and data
reduction techniques are described in \citet{henry99}.  Further details
on the development and operation of the automated telescopes can be found in
\citet{henry1995a,henry1995b} and \citet{eaton03}.

For 61~Vir, we used the two comparison stars HD~113415 (C1, $V=5.58$,
$B-V=0.56$, F7~V) and HD~114946 (C2, $V=5.33$, $B-V=0.87$, G8~III-IV).
The individual Str\"{o}mgren $b$ and $y$ differential magnitudes have
been corrected for differential extinction with nightly extinction
coefficients and transformed to the Str\"{o}mgren system with yearly mean
transformation coefficients.  Since 61~Vir lies at a declination of
$-18\arcdeg$, the photometric observations from Fairborn were made at
air mass 1.6\,--\,1.8.  Therefore, to maximize the precision of the measurements,
we combined the Str\"{o}mgren $b$ and $y$ differential magnitudes into
a single $(b+y)/2$ passband.  We also computed the differential magnitudes
of 61~Vir with respect to the mean brightness of the two comparison stars.
Because we are interested only in variability timescales of days to weeks, we
have normalized the final differential magnitudes so that the mean brightness
of each observing season is equal to zero.  This effectively removes any
long-term brightness variability in the two comparison stars as well as in
61~Vir.  (However, the standard deviation of the seasonal mean differential
magnitudes was only 0.00033~mag before normalization.)

\begin{figure}
\epsscale{0.9}
\plotone{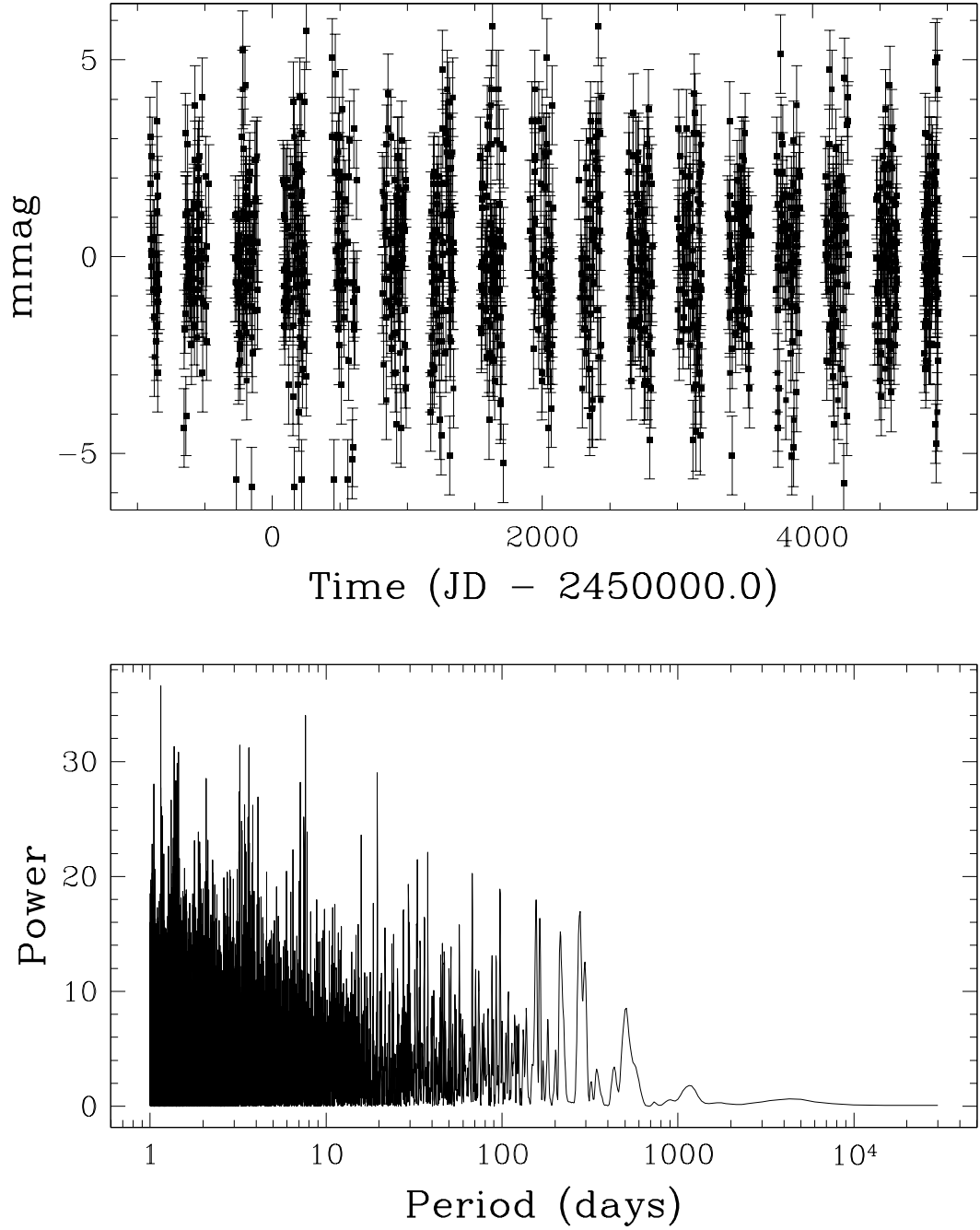}
\caption{Differential photometry of 61~Vir (top panel).  Periodogram of the
photometry (bottom panel).}
\label{photometry}
\end{figure}

A total of 1194 normalized differential magnitudes from 17 observing
seasons are plotted in the top panel of Figure~\ref{photometry}.  The data
scatter about their mean with a standard deviation $\sigma=0.00196$~mag,
which provides an upper limit to possible brightness variation in
61~Vir.  A Lomb-Scargle periodogram of the photometric measurements is
shown in the bottom panel of Figure~\ref{photometry} and reveals no
significant periodicities within the data.  We computed least-squares sine
fits for the three RV planet candidates described below; the
semi-amplitudes of the lightcurve fits were $0.00016 \pm 0.00007$,
$0.00011 \pm 0.00008$, and $0.00014 \pm 0.00007$ mag for the three periods
4.215, 38.012, and 123.98 days, respectively. We conclude that the
photometric constancy of 61~Vir supports planetary reflex motion as the
cause of the RV variations described in the next section.

\section{The Planetary System Orbiting 61~Vir}

\begin{figure}
\includegraphics[angle=-90,scale=0.65]{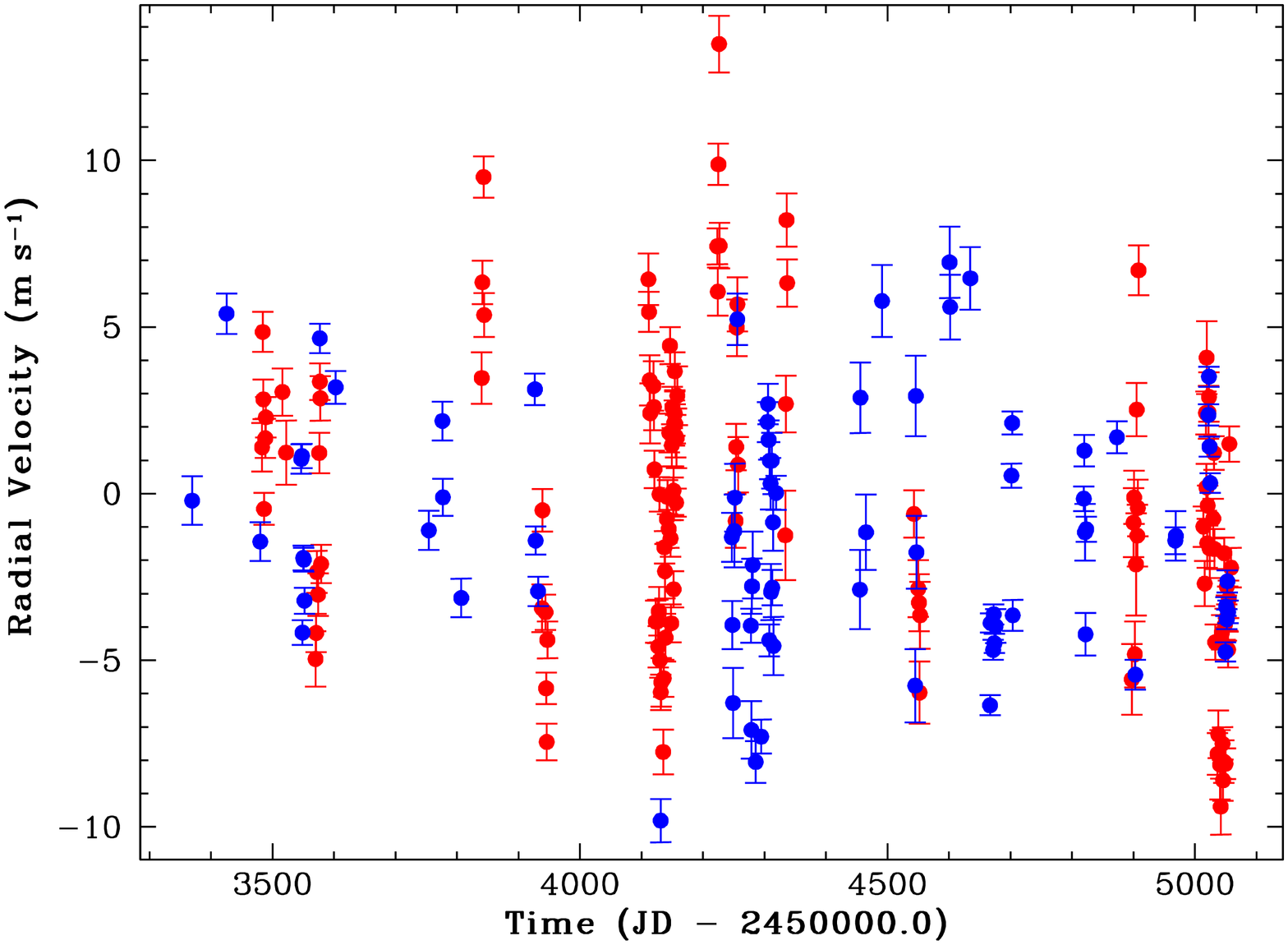}
\caption{Relative radial velocities of 61~Vir.  Velocities obtained at the AAT
are shown in red, and those from Keck are shown in blue.}
\label{velocities}
\end{figure}

The combined RV data from the AAPS and Keck telescopes show a
root-mean-square (rms) scatter of 4.1\,\ms\ about the mean velocity.  This
includes an offset of 0.895 \ms between the two telescopes (Keck - AAT) that 
was left as a free parameter and emerged from the three-planet Keplerian fit.
This rms significantly exceeds both
the scatter due to the underlying precisions of both the Keck and AAPS Doppler
measurement pipelines {\em and} the scatter expected in this star due to its
predicted 1.5\,\ms\ level of stellar jitter.
Figure~\ref{velocities} shows the combined RV data set.

\begin{figure}
\plotone{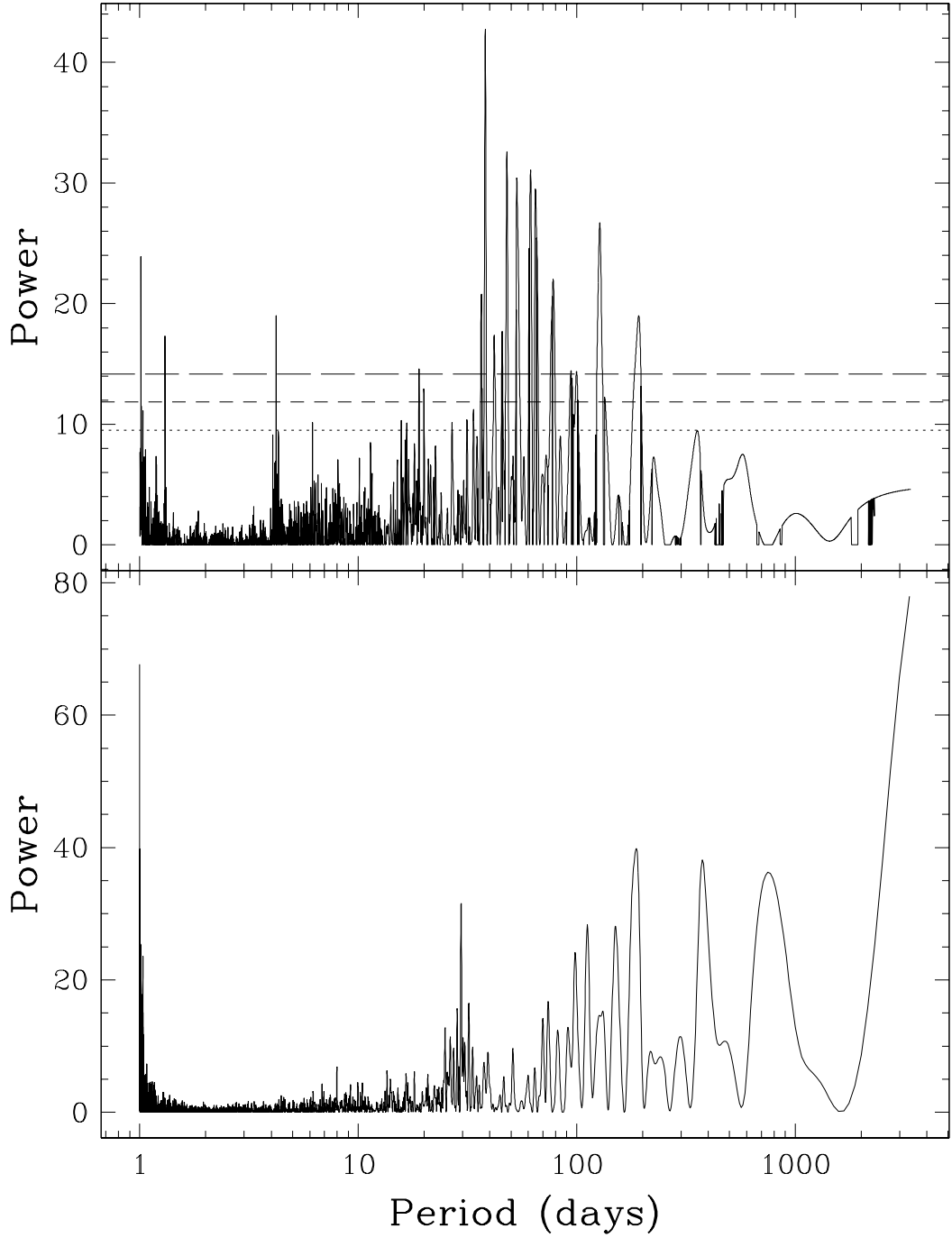}
\caption{Top panel: circular periodogram of the combined RV data
set for 61~Vir.  Bottom panel: power spectral window of the combined RV
data for 61~Vir.}
\label{period0}
\end{figure}

Figure \ref{period0} (top panel) shows the circular periodogram of the combined RV
data set. This is the power at any period associated with fitting
circular orbits to the data. Power at each sampled period is proportional to
the relative improvement in the fit quality for a circular model versus a
constant velocity model --- that is, the relative drop in $\chi_{\nu}^2$.  Figure \ref{period0}
(bottom panel) shows the spectral window or power spectrum due to the sampling times
\citep{dee75}. This spectral window illustrates spurious power that can appear in the data
merely from the sampling times alone. 

The periodogram of the RV data (top panel of Figure \ref{period0}) shows a
number of significant signals, with the strongest peak
occurring at a period of $P=38.13$ days. The false alarm probability (FAP) of
this peak is estimated (adopting the procedure described in \cite{cum04}) to be
FAP$\sim3.8\times10^{-16}$. The horizontal lines in Figure \ref{period0} (top
panel) and all similar figures below indicate (top to bottom) the 0.1\%, 1\%,
and 10\% FAP levels. This $P=38.13$-day signal, furthermore, lies far
from the periods favored by the sampling window (Figure~\ref{period0}; bottom
panel),  which produces spurious power at periods near 357, 191, 95.5, 652,
29.6, and 110 days. (Signals near these periods would naturally attract
suspicion as being artifacts of the lunar and yearly periods on telescope
scheduling.)

\begin{figure}
\plotone{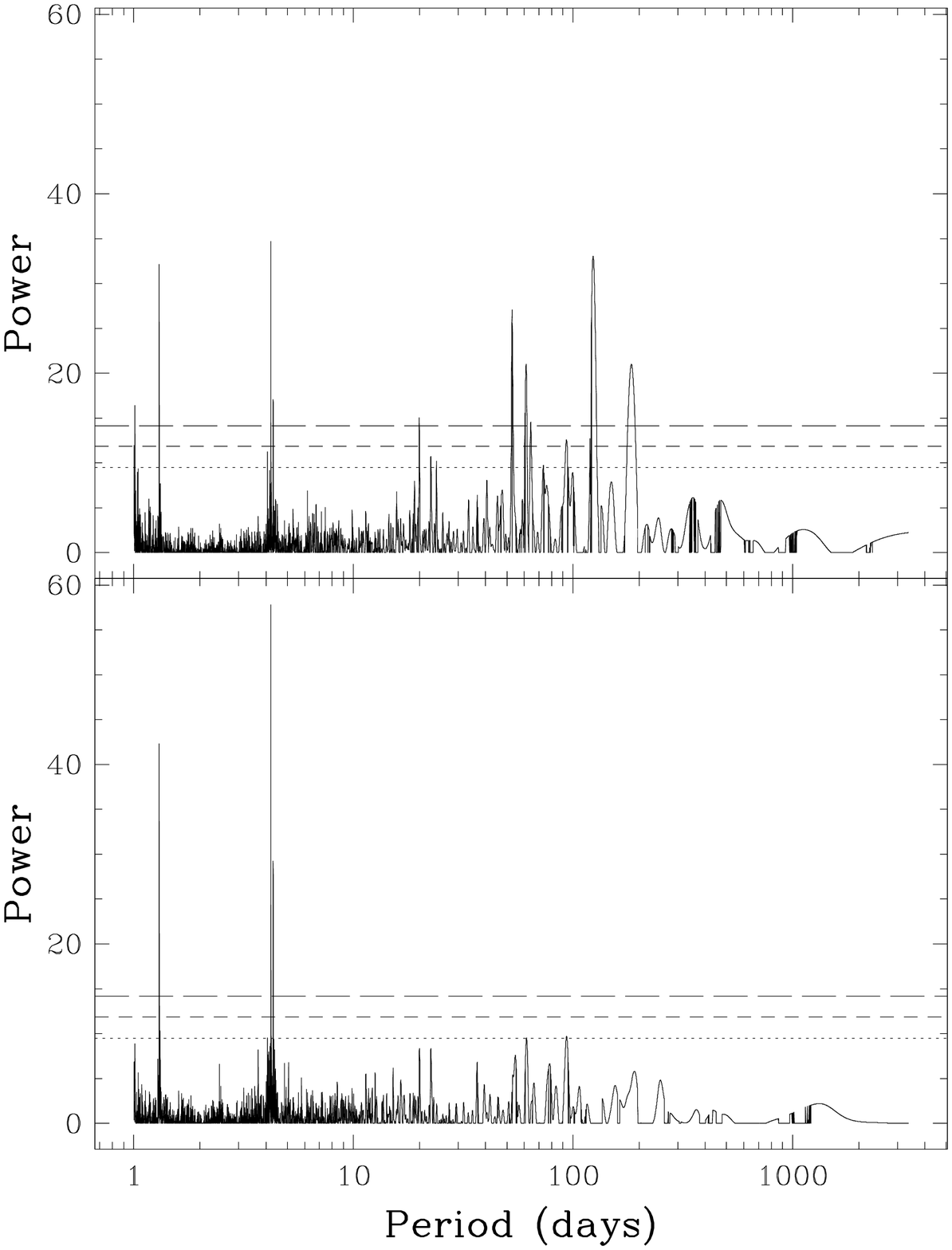}
\caption{Top panel: circular periodogram of the one-planet residuals of the
combined RV data set for 61~Vir.  Bottom panel: circular
periodogram of the two-planet residuals of the combined RV data set
for 61~Vir.}
\label{period12}
\end{figure}

The mass of the host star is assumed to be 0.95 $M_{\odot}$, the isochrone mass
of \cite{vaf05}. Based on the periodogram, we then
fit a planet of mass $M\sin{i}=15 M_{\oplus}$ and $P=38.20$ days on a circular
orbit to the RV data. The presence of this planet (with RV
semi-amplitude $K=2.98$ \ms) reduces the rms scatter of the velocity
residuals to 3.39 \ms. Figure~\ref{period12} (top panel) shows the
periodogram of the residuals to the one-planet fit which has strong peaks at
$P=124$ days and $P=4.21$ days.
The $P=124$-day signal has FAP$\sim6.3\times10^{-12}$ and can be modeled with a
companion with $K=3.36$ \ms and $M\sin{i}=24 M_{\oplus}$.  The addition of this
planet further reduces the rms scatter to 2.78 \ms.

Figure~\ref{period12} (bottom panel) shows the periodogram of the residuals to
the two-planet fit which has a highly significant peak at a period of $P=4.215$
days.  As a consequence of the fact that the outer two planets, once removed,
have significantly reduced the variance in the RV data, the FAP
of the 4.215-day signal in the residuals is very small,
FAP$\sim1.1\times10^{-22}$. We are thus confident that this periodicity is real,
and we ascribe it to the presence of a $P=4.2149$ days,
$M\sin{i}=5.1 M_{\oplus}$ companion.  There is another peak at 1.3 days that
arises as an alias of the 4.2-day periodicity beating with the nightly sampling.
The three-planet model has an rms scatter of 2.17 \ms.

Given the three-planet model, we can look either for solutions in which the
planetary orbits are circular or solutions where the eccentricities are allowed
to float. Inclusion of eccentricities provides only a modest improvement to the
orbital fit, and we conclude that a significant amount of additional Doppler
velocity monitoring will be required to improve measurement of the eccentricities.
For a detailed discussion of how additional measurements can reduce uncertainties
in orbital elements, see \cite{for05}.

In Tables~\ref{planetparamsnoe} and \ref{planetparamswithe}, we
present our best-fit versions of the system under the assumption of circular
orbits (Table~\ref{planetparamsnoe}) and with the additional degrees of freedom
provided by fully Keplerian trajectories (Table~\ref{planetparamswithe}). For
the orbital fits, we assume $i=90^{\circ}$ and $\Omega=0^{\circ}.$ We have
verified that the inclusion of planet-planet gravitational interactions in the
fit are unnecessary. Most of our modeling involves simply adding Keplerians. However, for cases where we expect significant gravitational interactions between companions, we carry out a more detailed modeling that involves using Hermite fourth order approximations to the planets' trajectories, accurately characterizing such gravitational interactions. If the Hermite fourth order calculations do not differ significantly from simple summed Keplerians, as is the case here, we conclude that planet-planet interactions are not necessary.

Uncertainties are based on 1000 bootstrap trials for which
we follow the procedure in Section 15.6 from \citet{pre92}. We take the
standard deviations of the fitted parameters to the bootstrapped RVs as the
uncertainties. The fitted mean anomalies are reported at epoch JD 2453369.166.
Our fitting was carried out with the publicly available {\it Systemic Console} \citep{systemic}.

\begin{figure}
\plotone{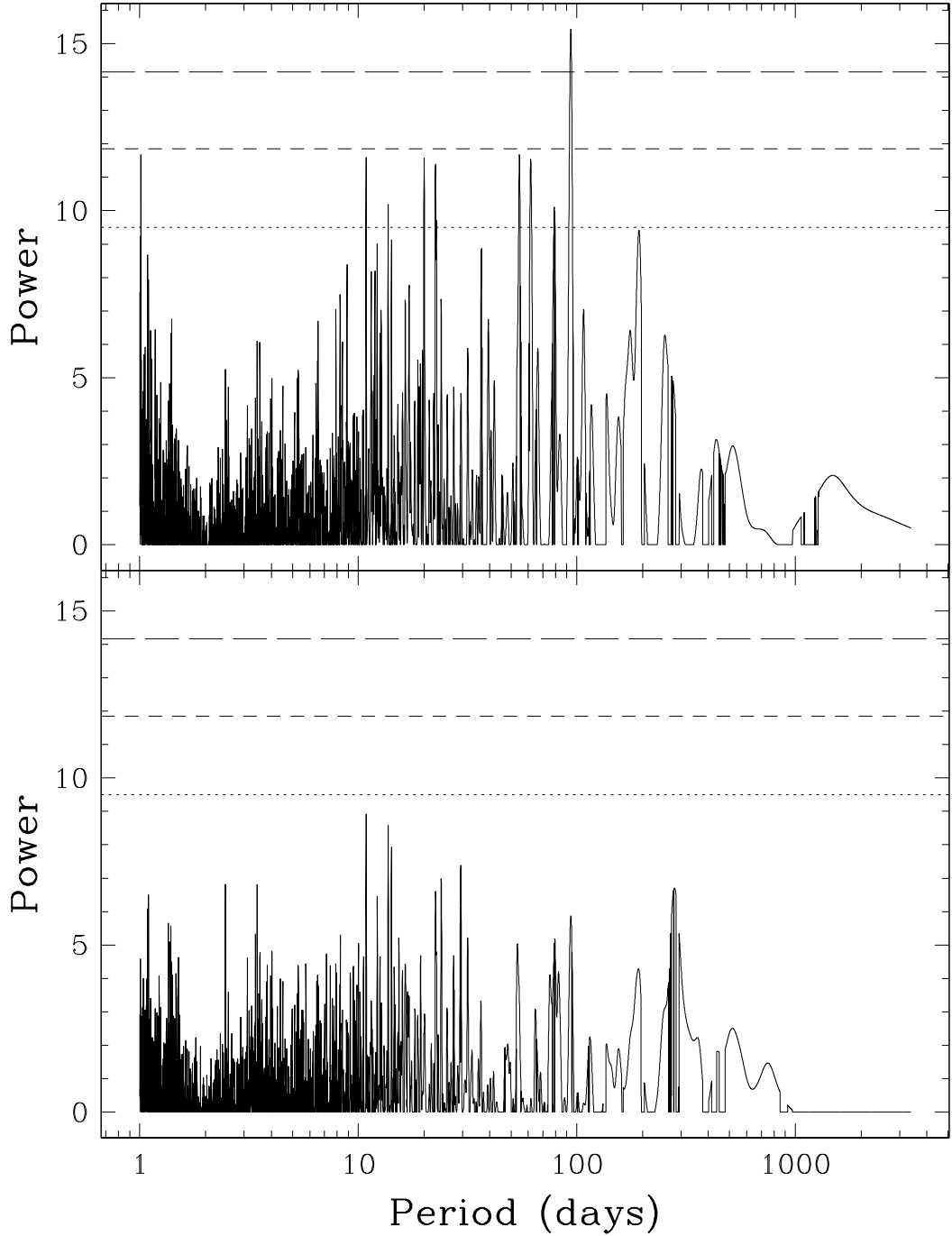}
\caption{Top panel: circular periodogram of the three-planet residuals, assuming
circular orbits, of the combined RV data set for 61~Vir.
Bottom panel: circular periodogram of the three-planet residuals, with fitted
eccentricities.}
\label{period3}
\end{figure}

Figure \ref{period3} (top panel) shows the power spectrum of the velocity
residuals for the three-planet $e=0$ fit.  There is a significant peak near
94 days with FAP$\sim0.0003$.  Although using this period in a four-planet circular
fit results in a significant improvement in $\chi_{\nu}^2$, the rms decrease
from 2.17 \ms for the three-planet model to 2.00 \ms is not significant.
Additionally, the fitted $K=1.44$ \ms is significantly smaller than the scatter
around the model.  But perhaps most significantly, this fourth peak almost
exactly corresponds to a peak in the window function of our data
(Figure~\ref{period0}; lower panel), making any association of it with a
real planet questionable.

The chi-squared of our three-planet circular fit is 13.03 and results in a fit with an
rms of 2.17 \ms and estimated stellar jitter of 2.06 \ms. That estimate of the
stellar jitter is the jitter required to bring the chi-squared of the fit down to 1.0.
Thus, if the stellar jitter is 2.06 \ms, our three-planet fit is essentially perfect.
The difference between a true stellar jitter of 2.06 \ms and our estimated value of 1.5 \ms
is negligible, given the accuracy of such jitter estimates. Moreover, there are likely
to be other planets in this system such as the 94-day signal in the top panel of 
Figure \ref{period3}. Adding more planets to the model would further reduce the
stellar jitter component of the fit. But the present data set does not support adding
another Keplerian to the model.

Figure \ref{period3} (bottom panel) also shows the power spectrum of the
velocity residuals for the eccentric three-planet fit.  The tallest peak is near
10.9 days, but it is not significant with a FAP near 20\%.  Also, note that the
rms for the eccentric three-planet fit of 2.09 \ms is not significantly different
from the 2.17 \ms for the circular three-planet fit.  At this point, all of these
concerns indicate that many more observations are needed both to constrain the
eccentricities of the first three planets and to pursue the prospect of a
potential fourth planet in the system.

\begin{figure}
\includegraphics[angle=-90,scale=0.65]{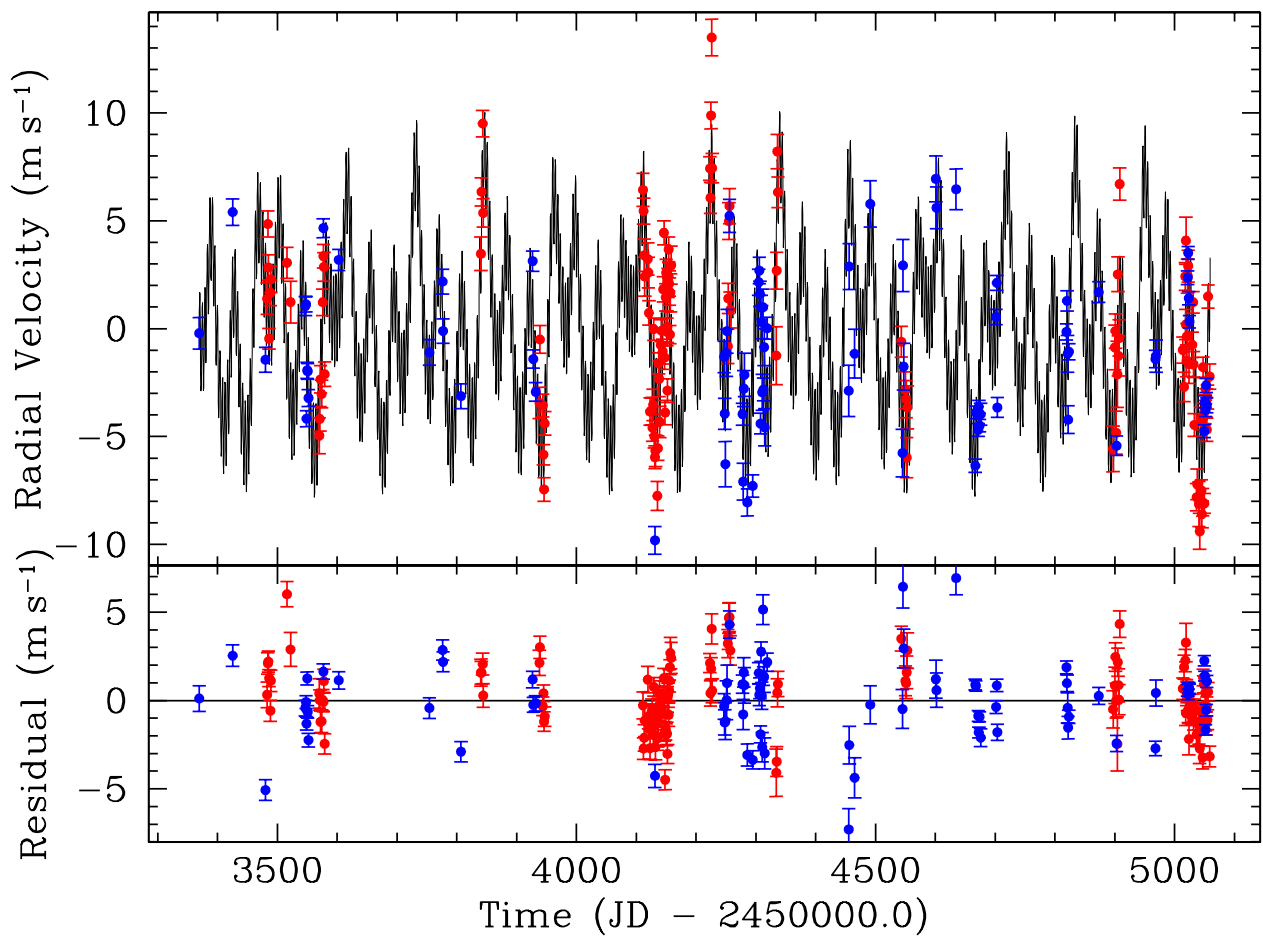}
\caption{Top panel: observed RVs of 61~Vir (red points for
AAT observations and blue for those from Keck) and model RV of
61~Vir due to the eccentric three-planet fit listed in Table~\ref{planetparamswithe}
(black curve).  The fitted relative offset of 0.895 \ms has been applied between
the two data sets from the two telescopes. Bottom panel: residual velocities
remaining after subtracting the eccentric three-planet model from the observations
(again, red is for AAT, and blue is for Keck).}
\label{model}
\end{figure}

\begin{figure}
\epsscale{0.75}
\plotone{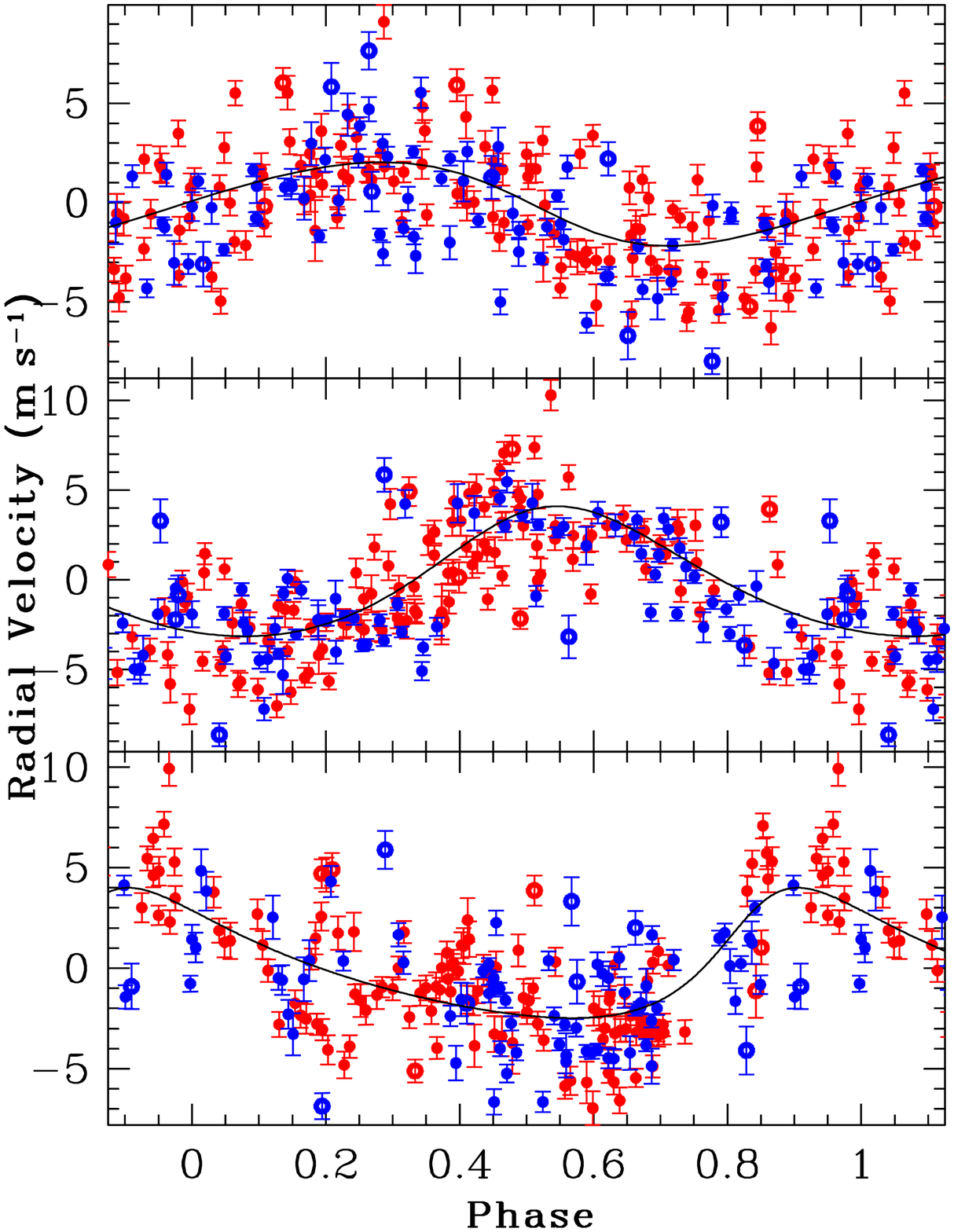}
\caption{Top panel: radial velocity of 61~Vir due to planet b folded at
4.215 days.
Center: radial velocity of 61~Vir due to planet c folded at
38.021 days.
Bottom panel: radial velocity of 61~Vir due to planet d folded at
123.01 days.
In each panel, the effect of the other two planets has been subtracted out.
The curves represent the model velocities due to each respective planet.
The AAT observations are shown in red, and the Keck velocities are
shown in blue. Open symbols represent observations that may be suspect
due to low S/N based on observing log notes.}
\label{phasedplot}
\end{figure}

To summarize up to this point, we have shown that the combined Keck and AAT RV
data show strong evidence for a three-planet system in orbit about the G5V star
61~Vir.  Figure~\ref{model} shows the barycentric RV of the star
computed with the eccentric three-planet model.  Figure~\ref{phasedplot} shows the
barycentric RV of the star computed with the eccentric model due
to each individual companion in the system. In each panel, the velocities are
folded at the period of each corresponding planet. Points marked as open symbols
denote observations that are suspect due to poor observing conditions that led to
low S/N in the spectra. All such points were previously noted in the observing logs
before any analysis occurred. Regardless of such suspicions, all observations were
used in all the analyses.

\section{Dynamical Analysis}

For any multiple-planet system, long-term orbital stability is a matter of both
interest and concern. Even precisely characterized, seemingly stable
configurations such as the Solar System are subject to chaotic planet-planet
interactions that can lead to orbit-crossing, collisions, and ejections on time
scales shorter than the stellar lifetime (for recent treatments see, e.g.,
\citet{lag08}; \citet{bat08}; \citet{lag09}). Given the uncertainties that
adhere to the orbital elements, the effort of generating a dynamical
characterization of a system such as 61~Vir b-c-d, is somewhat akin to producing
detailed maps of sand dunes. Nonetheless, it is useful to verify, via numerical
integration, whether the planetary configurations listed in
Tables~\ref{planetparamsnoe} and \ref{planetparamswithe} are dynamically stable.
Such an analysis is particularly useful in giving rough bounds on the allowed
coplanar inclinations relative to the line of sight to Earth. For simulations
that do not include the effects of tidal dissipation, we use fitted Newtonian
parameters for the initial states for long-term integrations.  With a time step
of 0.1 day, we use the MERCURY integration package \citep{cha99} for the
simulations.  We include the first order post-Newtonian term in the star's
potential as in \cite{lis01}.

If we assume the three orbits are initially circular, with periods, masses, and
mean anomalies given in Table~\ref{planetparamsnoe}, we find that the system is
stable for at least 365 Myr.  Additionally, assuming the system to be coplanar,
if we set the inclination to the sky plane to various values from $i=90^{\circ}$
all the way down to $i=1^{\circ}$ and perform a Newtonian fit for the other 11
parameters (three parameters per planet plus the two velocity offsets), $\chi^2_{\nu}$ does
not change significantly from the nominal $i=90^{\circ}$ fit.  We also find the
$i=1^{\circ}$ fit to result in a system that is stable for at least 50 Myr.  For
this inclination, the fitted masses exceed 1.0, 3.3, and 4.5 $M_{\rm Jup}$. This
system is stable because of the small eccentricities.  Thus, under the
assumption that the system is coplanar and the orbits are (nearly) circular, we
cannot place a lower bound on the inclination of the system.

\begin{figure}
\plotone{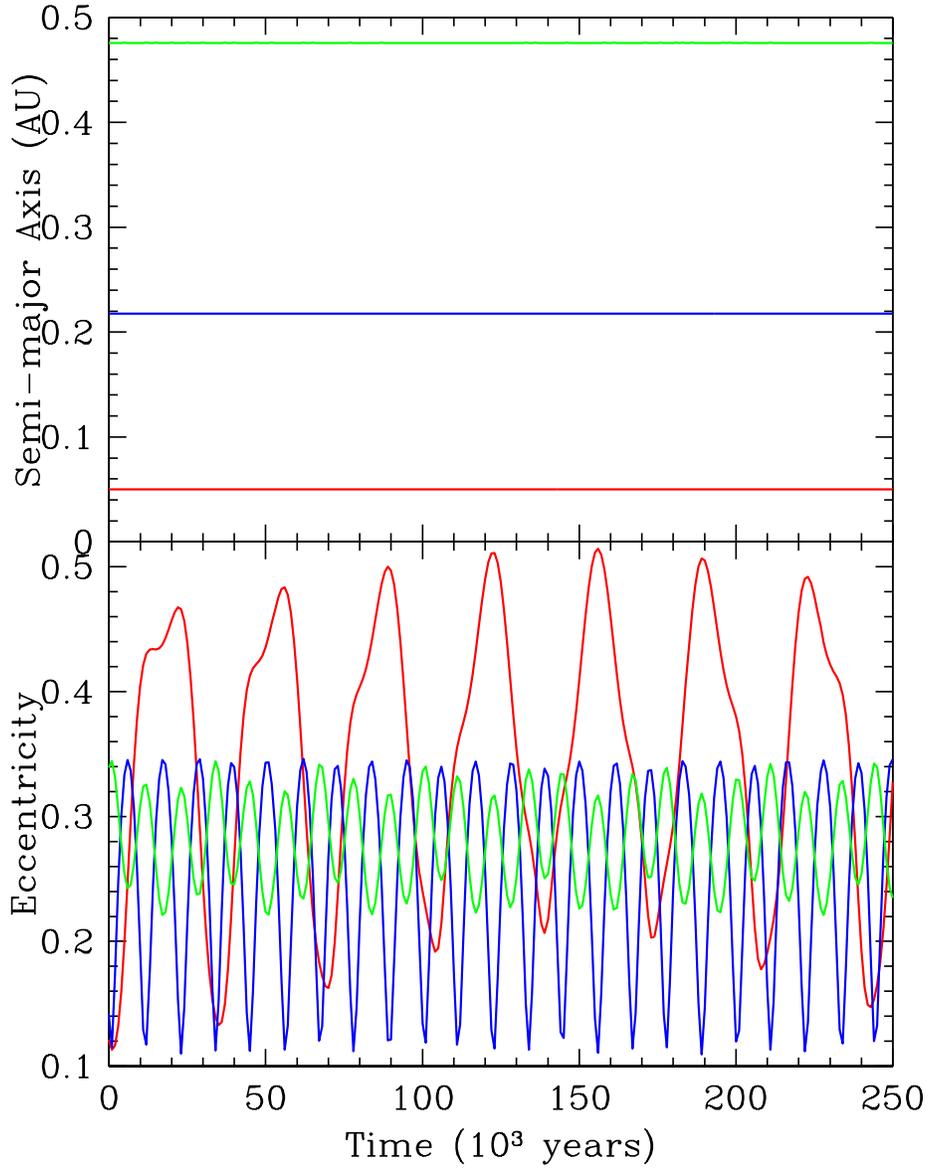}
\caption{Top panel: semi-major axes vs. time for the simulation using the
parameters from Table~\ref{planetparamswithe} as starting conditions.
Bottom panel: eccentricities vs. time for the same simulation.
For both panels, planet b is in red, c is in blue, and d is in green.}
\label{stability}
\end{figure}

The parameters of the floating-eccentricity version of the 61~Vir system given
in Table~\ref{planetparamswithe} were also used as the initial input conditions
for a $10^7$ yr simulation.  The three-planet configuration remained stable for
the full duration of this simulation.  Figure~\ref{stability} shows the
evolution of the planets' semi-major axes and eccentricities over the first
250,000 years of the simulation.  The planets interact with each other on
secular timescales, and the inner planet experiences a large excursion in
eccentricity (up to $e\sim$0.5) but is never disrupted.

With its period of 4.215 days, the innermost planet in the system is likely
subject to significant tidal dissipation in its interior. Tidal damping, in
fact, may have played an important role in the evolutionary history of this
multi-planet system. An investigation of this process requires a dynamical
theory that couples dissipative orbital evolution with planet-planet
gravitational perturbations among the individual system members. While the
coupling between dissipative and secular processes can be complex, there exists
an avenue toward insight regarding the overall dynamical state of the system.
In particular, it may be possible that the dissipative properties of the
innermost planet can be inferred. 

The key idea is that, under the influence of tidal dissipation, multiple-planet
systems approach stationary configurations. These so-called
{\it tidal fixed points} are discussed in the context of extrasolar planets by
\citet{wugo02} and more generally by \citet{mar07}. Fixed-point configurations
are characterized by either parallel or anti-parallel alignment of the apsides,
and simultaneous precession of the apsidal lines. Additionally, the
eccentricities of such systems are well determined and are not subject to
significant time variations.

Participation by the system in a fixed-point configuration hinges on the tidal
quality factor, $Q$, of planet b. Assuming that the system did not form in a
stationary configuration, and that the dissipation of planetary orbital energy
has been dominated during the past by tides raised on the innermost planet, it
can be shown that it takes $\sim 3$ circularization timescales, $\tau$, given
by \citet{yope81},
\begin{equation}
\tau = \frac{2}{21n} \left( \frac{Q}{k} \right) \left( \frac{m}{M_{\star}} \right) \left( \frac{a}{R} \right) ^{5},
\end{equation}
where $k$ is the planetary Love number and $R$ is the inner planet's radius, for
the system to become stationary \citep{mar07}. Therefore, assuming
$R = 2.45 R_{\oplus}$ and $k=0.34$, we expect that it would take
$\sim 700,000 \times Q$ years for the system to arrive at the fixed point. We
have confirmed this, by performing a numerical integration, using the tidal
friction formalism of \citet{egg98} and \citet{mar02}, and the eccentric fit
listed in Table~\ref{planetparamswithe} as initial conditions. Indeed, libration
of planet b around the fixed point begins after $\sim 850,000 \times Q$ years,
in rough agreement with theory. It is further interesting to note that the
best-fit eccentricities for planets c and d are in fact already very close to
the fixed-point values, and their perihelia are already set in a librating mode
at the beginning of the integration.

Thus, given the star's multi-billion year age, we can make the following claim:
if planet b has a Neptune-like $Q$ ($\sim30,000$), then the system is probably
not stationary. However, if the planet has a characteristic terrestrial
$Q$ ($\sim100$), then we can expect the system to be at a fixed point.  If the
system is at a fixed point, and the eccentricities are modest, we can use
modified Laplace-Lagrange secular theory to relate the planets' eccentricities
to each-other. The three-planet fixed point equations presented in \citet{bat09},
yield the relationships: $e_{c} \approx 5.11 e_{b}$, and
$e_{d} \approx 7.27 e_{b}$. Thus, by a precise determination of fixed-point
eccentricities, we can not only determine the dynamical state of the system,
but also begin to characterize the nature of planet b.

\section{Discussion}

We have detected three low-mass planets orbiting the nearby star 61~Vir. With a
RV semi-amplitude $K=2.09\pm0.23$ \ms, the
$M\sin{i}=5.10\,M_{\oplus}$ inner planet 61~Vir\,b is among the very
lowest amplitude companions yet detected using the Doppler velocity technique.
To date, the only announced planets with {\it smaller} amplitudes have been
Gliese~581e, with $K=1.85\pm0.23$ \ms \citep{mayor581e}, and HD~40307b, with
$K=1.97\pm0.11$ \ms \citep{mayor40307}. The inner planet in the HD~69830 system,
with $K=2.20\pm0.20$ \ms \citep{lov06} presents a slightly larger $K$.
Interestingly, all four of the lowest amplitude planets known are members of
multiple-planet systems containing three or more Neptune or super-Earth class
objects on orbits that would be ascribed to the inner zone of the terrestrial
region were they in our own solar system.

Given the observed multiplicity, negligible jitter, and extraordinary
photometric stability associated with the 61~Vir planetary system, it is
natural to speculate on the existence and detectability of additional planets.
As 61~Vir is a naked-eye visible solar-type star observable from both
hemispheres, it has been the target of many planet-search efforts in the past
25 years.  One of the first precision RV searches for planets, by
\citet{walker95}, observed 61~Vir for approximately 12 years. They present a
figure indicating rough upper limits of 1\,--\,2 Jupiter masses in the period range
between one and ten years.  \citet{cumming99} included 61~Vir in their
determination of companion limits from 11~yr of the Lick planet search.
They computed a 99\% upper limit on the velocity amplitude $K=19$ \ms\ and noted
a possible candidate companion with a period of 9.8~yr and $K\sim15$ \ms.
Later, \citet{Wit06} combined 23 years of velocities for 61~Vir, including
those data from \citet{walker95} and from the McDonald Observatory long-term
planet search program, to derive companion-mass upper limits of
$1.16\,M_{\rm Jup}$ at 3~AU and $1.69\,M_{\rm Jup}$ at 5.2~AU.  From these
studies, and the lack of any significant trends in the RV data, it
would appear that there are no planets in the 61~Vir system with masses
substantially greater than that of Jupiter.  Undetected sub-Jovian mass planets
may be present in orbits with periods longer than the extant high-precision
velocity data.

A further interesting clue to the nature of the 61~Vir system was recently
provided by \citet{tanner09}, who detected an infrared excess at $160\mu$ using
\textit{Spitzer} observations. The excess $160\mu$ flux implies the presence of
a dust disk that is continuously replenished by collisions within a cold
Kuiper-Belt-like disk of planetesimals surrounding the star.  \citet{tanner09}
also suggest that the disk is resolved at $70\mu$ (G. Bryden et al.\ 2009, in
preparation). Assuming that the emitting grains are black bodies, the disk spans
an annulus from $R_{\rm inner}=96\pm$5~AU to $R_{\rm outer}=195\pm$10~AU, with the dust
temperature at 55-45~K (inner to outer). An improved model of the disk, however,
can be achieved by assuming the emission arises from silicate grains with size
$0.25\mu$.  In this case, the disk spans $R_{\rm inner}=120\pm$20~AU to
$R_{\rm outer}=220\pm$10~AU with a dust temperature of 24-19~K. An excess at $70\mu$
was also detected with \textit{Spitzer} by \citet{trilling08}.

The 61~Vir system is consistent with the population of planets postulated by
\citet{mayor40307}, who inferred that fully one-third of solar-type stars in the
immediate galactic neighborhood are accompanied by Neptune- (or lower) mass
planets having orbital periods of 50 days or less. This putative population was
largely unexpected, and has spurred a good deal of recent theoretical work
geared to explain its existence. We note that the habitable zone of 61~Vir is
located in the vicinity of a 300-day orbit. At this separation from the parent
star, a $2\,M_{\oplus}$ planet induces a RV semi-amplitude $K=0.2$
\ms.  Assuming a combined stellar jitter and median internal uncertainty of 1.5
\ms, this is a factor of about 30 greater than the 0.05 \ms precision required to
make a $4\sigma$ measurement of a such a small signal. If precision scales as the
square root of the total number of Doppler measurements, a total of $30^{2}$ or 900
velocities (an additional 700 more) would allow for the detection of such a planet with
a S/N of 4.

With a period of only 4.215 days ($a=$\,0.05~AU), 61~Vir\,b may transit its
star.  The \textit{a priori} transit probability is $\sim9$\% (assuming
$R_{\star}=1\,R_{\odot}$).  Due to the low mass of 61~Vir\,b, a successful
transit observation would allow a detailed physical probe of a fundamentally new
class of planet. Currently,  the lowest mass transiting planet is CoRoT-7b, with
a radius of $1.68\pm0.09\,R_{\oplus}$ \citep{corot7}.  CoRoT-7b orbits a G9V
star and has a transit depth of only $\Delta F/F=3.4\times 10^{-4}$. Its mass is
estimated by Queloz et al. (2009) to be $4.8\pm0.8M_{\oplus}$. By comparison,
if 61 Vir~b is transiting, sin $i~\sim 1$, and hence it would have a mass of
$5.10\,M_{\oplus}$. Assuming
that the planet migrated inward from beyond the ice line of 61~Vir's
protoplanetary disk, it is likely made mostly of water. The models of
\citet{fortney07} indicate a $2.45\,R_{\oplus}$ radius for such a planet,
leading to a central transit depth of $\sim0.54$ millimagnitudes. Such a transit
could be detected from space using, for example, the Warm Spitzer platform.

Of the currently known planets discovered by the RV technique with $M\sin{i} < 10 M_{\oplus}$ (10 planets;
6 host stars), half orbit M dwarf stars. Those are GJ 581c,d,e
\citep{mayor581e}, GJ 876d \citep{rivera05}, and GJ 176b \citep{forveille09}.
The rest of those planets orbit K stars: HD 7924b \citep{howard09}, HD 181433b
\citep{bouchy09}, and HD 40307b,c,d \citep{mayor40307}. This is of course a
selection effect, since late-type stars have lower masses and hence a planet of
a given mass would produce a larger RV signal. 61~Vir, therefore,
is a unique new system in that not only does it host multiple low-mass planets,
but also it is the first G-type Sun-like star found through the RV technique to host a planet with
$M\sin{i} < 10 M_{\oplus}$.

The detection of the three low-$K$ planets reported here around this previously
well-studied, bright, nearby star was made possible by the combined cadence and
high precision of the AAPS and Keck surveys. As cadence and time bases grow in
the RV monitoring of chromospherically quiet, nearby stars, complex
planetary systems, like 55~Cnc, and now 61~Vir, are becoming increasingly
common.  This is only our first reconnaissance of this fascinating and quite
nearby system.  As more RV data are collected, the orbital
ephemerides of the planets will become better determined, and more planets will
probably be revealed. 

The 61~Vir system joins a growing class of exoplanet systems that have multiple
planets orbiting with periods less than an Earth-year.  Other examples are
HD~75732 (55~Cnc), HD~69830, GJ~581, HD~40307, and GJ~876.  The increasing
frequency of such systems portends that space-based transit surveys such as
COROT and KEPLER will find many multi-transiting systems.

\acknowledgments

We gratefully acknowledge the major contributions of fellow members of our
previous California-Carnegie Exoplanet team: Geoff Marcy, Jason Wright, Debra Fischer,
and Katie Peek in helping to obtain the RVs
presented in this paper.  S.S.V. gratefully acknowledges support from NSF grant
AST-0307493.  R.P.B. gratefully acknowledges support from NASA OSS Grant
NNX07AR40G, the NASA Keck PI program, and the Carnegie Institution of
Washington.  G.L. acknowledges support from NSF AST-0449986.  G.W.H. acknowledges
support from NASA, NSF, Tennessee State University, and the state of Tennessee
through its Centers of Excellence program.  The work herein is based on
observations obtained at the W. M. Keck Observatory, which is operated jointly
by the University of California and the California Institute of Technology,
and we thank the UC-Keck and NASA-Keck Time Assignment Committees for their
support.  We gratefully acknowledge the UK and Australian government support of
the AAT through their PPARC, STFC, DETYA, and DIISR
funding. C.G.T. and H.R.A.J. acknowledge support from STFC grant PP/C000552/1, while
C.G.T. acknowledges support from ARC Grant DP0774000. Travel support to the AAT has
been generously provided by the Anglo-Australian Observatory (to C.G.T., B.D.C., and
J.B.).  We are grateful for the extraordinary support we have received from the
AAT technical staff at the AAT --- E. Penny, R. Paterson, D. Stafford, F. Freeman,
S. Lee, J. Pogson, S. James, J. Stevenson, K. Fiegert, and W. Campbell.  We also
wish to extend our special thanks to those of Hawaiian ancestry on whose sacred
mountain of Mauna Kea we are privileged to be guests.  Without their generous
hospitality, the Keck observations presented herein would not have been
possible.  This research has made use of the SIMBAD database, operated at CDS,
Strasbourg, France.

{\it Facilities:} \facility{AAT (UCLES),} \facility{Keck (HIRES)}.

\begin{deluxetable}{lrr}
\tabletypesize{\scriptsize}
\tablecolumns{3}
\tablewidth{0pt}
\tablecaption{Stellar Parameters for 61~Vir}
\tablehead{
\colhead{Parameter} & \colhead{Value} & \colhead{Reference}\\
}
\startdata
\label{stellarparams}
Spec.~Type          & G5V                       & \citet{cen07} \\
Mass ($M_{\odot}$)  & 0.95$^{+0.04}_{-0.03}$    & \citet{vaf05} \\
                    & 0.942$^{+0.034}_{-0.029}$ & \citet{takeda07} \\
Radius ($R_{\odot}$) & 0.963$\pm$0.011          & \citet{vaf05} \\
                    & 0.98$\pm$0.03             & \citet{takeda07} \\
Luminosity ($L_{\odot}$) & 0.805$\pm$0.028       & \citet{vaf05} \\
                        & 0.804$\pm$0.005       & \citet{sou08} \\
Distance (pc)       & 8.52$\pm$0.05             & \citet{per97} \\
$V\sin{i}$ (\kms)   & 2.2                       & \citet{vaf05} \\
                    & 1.9                       & \citet{desidera06} \\
$\log{R'_{\rm HK}}$ & -4.93                     & \citet{hall07} \\
                    & -4.96                     & \citet{henry96} \\
                    & -5.03                     & \citet{Wit06} \\
                    & -4.95                     & This work \\
$P_{\rm rot}$ (days) & 29                        & \citet{baliunas96} \\
Age (Gyr)           & 6.3$^{+3.3}_{-3.1}$        & \citet{vaf05} \\
                    & 8.96$^{+2.76}_{-3.08}$     & \citet{takeda07} \\
$[\rm Fe/H]$        & 0.05                      & \citet{vaf05} \\
                    & -0.01                     & \citet{cen07} \\
$T_{\rm eff}$ (K)   & 5571                      & \citet{vaf05} \\
                    & 5531                      & \citet{cen07} \\
                    & 5577$\pm$33               & \citet{ecuvillon06} \\
$\log{g}$           & 4.47                      & \citet{vaf05} \\
                    & 4.31                      & \citet{cen07} \\
                    & 4.45$^{+0.04}_{-0.03}$    & \citet{takeda07} \\
                    & 4.34$\pm$0.03             & \citet{ecuvillon06} \\
\enddata
\end{deluxetable}

\clearpage
\begin{deluxetable}{rrrr}
\tablecaption{Radial Velocities for 61~Vir}
\tablewidth{0pt}
\tablehead{
JD & RV & Error & Observatory \\
(-2450000)   &  (\ms) & (\ms) & 
}
\startdata
\label{115617vels}
3482.96440 &   2.10 &  0.73 & A\\
3483.96117 &   5.56 &  0.60 & A\\
3485.07996 &   3.54 &  0.59 & A\\
3486.04514 &   0.25 &  0.48 & A\\
3488.13906 &   2.37 &  0.59 & A\\
3489.08696 &   3.00 &  0.61 & A\\
3516.04043 &   3.76 &  0.71 & A\\
3521.94298 &   1.94 &  0.96 & A\\
3569.90141 &  -4.25 &  0.83 & A\\
3570.94617 &  -3.47 &  0.57 & A\\
3571.94048 &  -1.64 &  0.63 & A\\
3573.87030 &  -2.32 &  0.64 & A\\
3575.87530 &   1.93 &  0.61 & A\\
3576.89023 &   4.07 &  0.55 & A\\
3577.86776 &   3.57 &  0.67 & A\\
3578.91377 &  -1.40 &  0.57 & A\\
3840.21603 &   4.18 &  0.77 & A\\
3841.16188 &   7.05 &  0.65 & A\\
3843.13891 &  10.21 &  0.62 & A\\
3844.08319 &   6.07 &  0.66 & A\\
3937.93317 &  -2.73 &  0.72 & A\\
3938.93436 &   0.21 &  0.64 & A\\
3943.88525 &  -2.85 &  0.53 & A\\
3944.88569 &  -5.13 &  0.47 & A\\
3945.88638 &  -6.74 &  0.55 & A\\
3946.88960 &  -3.68 &  0.55 & A\\
4111.21051 &   7.14 &  0.77 & A\\
4112.21939 &   6.16 &  0.60 & A\\
4113.22990 &   4.11 &  0.75 & A\\
4114.26145 &   3.12 &  0.90 & A\\
4119.24597 &   3.94 &  0.74 & A\\
4120.22002 &   3.31 &  0.70 & A\\
4121.21210 &   1.44 &  0.56 & A\\
4123.23541 &  -3.13 &  0.63 & A\\
4126.19788 &  -3.12 &  0.65 & A\\
4127.20219 &  -3.88 &  0.62 & A\\
4128.20519 &  -2.82 &  0.73 & A\\
4129.20022 &   0.69 &  0.52 & A\\
4130.19130 &  -4.28 &  0.53 & A\\
4131.20082 &  -5.25 &  0.53 & A\\
4132.20696 &  -4.95 &  0.73 & A\\
4135.19894 &  -7.04 &  0.67 & A\\
4136.21567 &  -4.83 &  0.56 & A\\
4137.21358 &  -0.90 &  0.49 & A\\
4138.19311 &  -1.62 &  0.82 & A\\
4139.18689 &  -3.61 &  0.72 & A\\
4141.21764 &  -0.05 &  0.71 & A\\
4142.20644 &   0.61 &  0.51 & A\\
4144.14374 &  -0.35 &  0.57 & A\\
4145.17237 &   2.53 &  0.58 & A\\
4146.19289 &   5.15 &  0.55 & A\\
4147.21096 &  -0.63 &  0.55 & A\\
4148.24103 &  -3.18 &  0.57 & A\\
4149.18109 &   2.16 &  0.63 & A\\
4150.21285 &   3.30 &  0.62 & A\\
4151.22296 &   0.56 &  0.64 & A\\
4152.09352 &   0.80 &  0.73 & A\\
4152.25396 &  -2.16 &  0.54 & A\\
4153.17187 &   2.83 &  0.68 & A\\
4154.09474 &   4.38 &  0.57 & A\\
4154.27050 &   3.11 &  0.66 & A\\
4155.06551 &   2.80 &  0.59 & A\\
4155.27322 &   2.43 &  0.64 & A\\
4156.18613 &   0.44 &  0.42 & A\\
4157.16889 &   2.37 &  0.89 & A\\
4158.18706 &   3.65 &  0.88 & A\\
4223.12188 &   8.13 &  0.54 & A\\
4224.15218 &   6.77 &  0.72 & A\\
4225.08458 &  10.59 &  0.62 & A\\
4226.02062 &  14.20 &  0.85 & A\\
4227.02043 &   8.15 &  0.68 & A\\
4252.97759 &  -0.11 &  0.80 & A\\
4254.01661 &   2.11 &  0.70 & A\\
4254.91868 &   5.69 &  0.85 & A\\
4255.98460 &   6.39 &  0.81 & A\\
4257.07180 &   1.58 &  0.83 & A\\
4333.86604 &  -0.54 &  1.34 & A\\
4334.86018 &   3.40 &  0.85 & A\\
4335.85304 &   8.92 &  0.80 & A\\
4336.84663 &   7.03 &  0.71 & A\\
4543.06967 &   0.10 &  0.71 & A\\
4550.11293 &  -2.14 &  0.85 & A\\
4551.09816 &  -2.56 &  0.86 & A\\
4552.14240 &  -5.26 &  0.93 & A\\
4553.10980 &  -2.94 &  1.00 & A\\
4897.20941 &  -4.86 &  1.06 & A\\
4900.19861 &  -0.16 &  1.03 & A\\
4901.16839 &   0.60 &  0.80 & A\\
4902.21654 &  -4.11 &  0.99 & A\\
4904.20346 &  -1.41 &  1.53 & A\\
4905.27335 &   3.23 &  0.80 & A\\
4906.22330 &  -0.55 &  0.93 & A\\
4907.21266 &   0.28 &  0.85 & A\\
4908.21531 &   7.41 &  0.75 & A\\
5013.83842 &  -0.28 &  0.61 & A\\
5015.83950 &  -1.99 &  0.68 & A\\
5017.84841 &   3.13 &  0.64 & A\\
5018.90225 &   0.90 &  0.70 & A\\
5018.95764 &   4.79 &  1.09 & A\\
5019.99765 &  -0.78 &  0.73 & A\\
5020.83414 &   0.35 &  0.60 & A\\
5021.88251 &   3.15 &  0.79 & A\\
5022.89587 &   3.61 &  0.74 & A\\
5023.87589 &  -0.93 &  0.89 & A\\
5029.86175 &  -0.04 &  0.59 & A\\
5030.97802 &   1.93 &  0.50 & A\\
5031.84137 &  -0.95 &  0.60 & A\\
5032.92564 &  -3.75 &  0.53 & A\\
5036.86026 &  -7.10 &  0.63 & A\\
5037.84849 &  -6.51 &  0.71 & A\\
5040.85306 &  -7.42 &  1.04 & A\\
5041.95419 &  -8.68 &  0.84 & A\\
5043.84520 &  -3.45 &  0.52 & A\\
5044.84224 &  -6.79 &  0.49 & A\\
5045.83918 &  -7.89 &  0.62 & A\\
5046.91766 &  -7.33 &  0.64 & A\\
5047.86724 &  -1.08 &  0.47 & A\\
5048.85263 &  -3.21 &  0.47 & A\\
5049.86609 &  -7.39 &  0.46 & A\\
5050.87397 &  -3.02 &  0.68 & A\\
5051.86595 &  -2.08 &  0.53 & A\\
5052.88848 &  -2.74 &  0.66 & A\\
5053.88182 &  -3.97 &  0.54 & A\\
5054.86448 &  -2.46 &  0.56 & A\\
5055.85374 &   2.20 &  0.53 & A\\
5058.89140 &  -1.51 &  0.59 & A\\
3369.16663 &   1.69 &  0.73 & K\\
3425.07667 &   7.30 &  0.61 & K\\
3479.96032 &   0.46 &  0.58 & K\\
3546.78158 &   2.94 &  0.44 & K\\
3547.76796 &   3.03 &  0.36 & K\\
3548.80045 &  -2.27 &  0.37 & K\\
3549.80104 &  -0.03 &  0.37 & K\\
3550.82365 &  -0.08 &  0.36 & K\\
3551.80855 &  -1.31 &  0.39 & K\\
3576.75813 &   6.56 &  0.44 & K\\
3602.72508 &   5.09 &  0.49 & K\\
3754.09220 &   0.80 &  0.59 & K\\
3776.17013 &   4.08 &  0.58 & K\\
3777.08603 &   1.79 &  0.56 & K\\
3806.94545 &  -1.23 &  0.58 & K\\
3926.74769 &   5.03 &  0.47 & K\\
3927.78252 &   0.49 &  0.42 & K\\
3931.82248 &  -1.03 &  0.44 & K\\
4131.14294 &  -7.91 &  0.65 & K\\
4246.81083 &   0.59 &  0.73 & K\\
4247.94552 &  -2.04 &  0.72 & K\\
4248.81966 &  -4.38 &  1.05 & K\\
4250.80828 &   0.78 &  1.09 & K\\
4251.81149 &   1.78 &  1.02 & K\\
4255.75898 &   7.13 &  0.77 & K\\
4277.73265 &  -2.06 &  0.51 & K\\
4278.73377 &  -5.19 &  0.86 & K\\
4279.73075 &  -0.88 &  0.83 & K\\
4280.74026 &  -0.24 &  1.00 & K\\
4285.76472 &  -6.15 &  0.63 & K\\
4294.73843 &  -5.39 &  0.51 & K\\
4304.73624 &   4.05 &  0.60 & K\\
4305.73658 &   4.59 &  0.61 & K\\
4306.80060 &   3.51 &  0.59 & K\\
4307.73485 &  -2.50 &  0.48 & K\\
4308.73551 &   2.89 &  0.56 & K\\
4309.73851 &   2.20 &  0.77 & K\\
4310.73545 &  -1.05 &  0.84 & K\\
4311.73296 &   2.89 &  0.84 & K\\
4312.73423 &  -0.92 &  0.53 & K\\
4313.73496 &   1.04 &  0.85 & K\\
4314.73608 &  -2.67 &  0.87 & K\\
4318.77353 &   1.92 &  0.51 & K\\
4455.16774 &  -0.98 &  1.19 & K\\
4456.15784 &   4.78 &  1.06 & K\\
4465.13890 &   0.74 &  1.13 & K\\
4491.10891 &   7.68 &  1.08 & K\\
4545.03961 &  -3.86 &  1.10 & K\\
4546.03027 &   4.83 &  1.21 & K\\
4547.03936 &   0.14 &  1.09 & K\\
4600.92698 &   8.84 &  1.07 & K\\
4601.87647 &   7.50 &  0.97 & K\\
4634.78181 &   8.36 &  0.94 & K\\
4666.78904 &  -4.45 &  0.30 & K\\
4667.78945 &  -1.98 &  0.29 & K\\
4671.80264 &  -2.79 &  0.30 & K\\
4672.78558 &  -1.72 &  0.30 & K\\
4673.79754 &  -2.59 &  0.29 & K\\
4675.79484 &  -2.07 &  0.51 & K\\
4701.73724 &   2.44 &  0.36 & K\\
4702.73348 &   4.02 &  0.35 & K\\
4703.73110 &  -1.75 &  0.46 & K\\
4819.16638 &   1.75 &  0.37 & K\\
4820.17627 &   3.19 &  0.47 & K\\
4821.14761 &   0.74 &  0.85 & K\\
4822.14603 &  -2.32 &  0.64 & K\\
4823.16898 &   0.83 &  0.38 & K\\
4873.10516 &   3.59 &  0.48 & K\\
4903.14262 &  -3.53 &  0.45 & K\\
4967.98451 &   0.49 &  0.40 & K\\
4968.97445 &   0.63 &  0.74 & K\\
5021.83208 &   4.26 &  0.32 & K\\
5022.84171 &   5.41 &  0.30 & K\\
5023.74658 &   3.31 &  0.30 & K\\
5024.84232 &   2.22 &  0.29 & K\\
5049.76443 &  -2.85 &  0.29 & K\\
5050.76541 &  -1.49 &  0.28 & K\\
5051.75441 &  -1.87 &  0.30 & K\\
5052.75467 &  -0.73 &  0.33 & K\\
5053.76245 &  -1.65 &  0.34 & K\\
\enddata
\end{deluxetable}

\begin{deluxetable}{lr@{$\pm$}lr@{$\pm$}lccr@{$\pm$}lr@{$\pm$}lr@{$\pm$}l}
\tabletypesize{\scriptsize}
\tablecolumns{10}
\tablewidth{0pt}
\tablecaption{Circular Orbit Solutions (Epoch JD 2453369.166)}
\tablehead{
\colhead{Planet} & \multicolumn{2}{c}{Period} & \multicolumn{2}{c}{$M$}       & \multicolumn{1}{c}{$e$} & \multicolumn{1}{c}{$\omega$}  & \multicolumn{2}{c}{K}     & \multicolumn{2}{c}{$M\sin{i}$}    & \multicolumn{2}{c}{$a$} \\
\colhead{}       & \multicolumn{2}{c}{(days)} & \multicolumn{2}{c}{(degrees)} & \multicolumn{1}{c}{}    & \multicolumn{1}{c}{(degrees)} & \multicolumn{2}{c}{(\ms)} & \multicolumn{2}{c}{($M_{\oplus}$)} & \multicolumn{2}{c}{(AU)}
}
\startdata
\label{planetparamsnoe}   
61 Vir b & 4.2149 & 0.0006 & 268 & 13 & 0 & n/a & 2.09 & 0.23 & 5.1  & 0.6  & 0.050201 & 0.000005 \\
61 Vir c & 38.012 & 0.036  & 210 & 16 & 0 & n/a & 3.58 & 0.25 & 18.2 & 1.3  & 0.2175   & 0.0001   \\
61 Vir d & 123.98 & 0.40   &  77 & 33 & 0 & n/a & 3.18 & 0.29 & 24.0 & 2.2  & 0.478    & 0.001    \\
\enddata
\end{deluxetable}

\begin{deluxetable}{lr@{$\pm$}lr@{$\pm$}lr@{$\pm$}lr@{$\pm$}lr@{$\pm$}lr@{$\pm$}
lr@{$\pm$}l}
\tabletypesize{\scriptsize}
\tablecolumns{10}
\tablewidth{0pt}
\tablecaption{Keplerian Orbital Solutions (Epoch JD 2453369.166)}
\tablehead{
\colhead{Planet} & \multicolumn{2}{c}{Period} & \multicolumn{2}{c}{$M$}
&
\multicolumn{2}{c}{$e$} & \multicolumn{2}{c}{$\omega$} &
\multicolumn{2}{c}{K} & \multicolumn{2}{c}{$M\sin{i}$} &
\multicolumn{2}{c}{$a$} \\
\colhead{} & \multicolumn{2}{c}{(days)} & \multicolumn{2}{c}{(degrees)}
&
\multicolumn{2}{c}{} &
\multicolumn{2}{c}{(degrees)} & \multicolumn{2}{c}{(\ms)} &
\multicolumn{2}{c}{($M_{\oplus}$)} & \multicolumn{2}{c}{(AU)}
}
\startdata
\label{planetparamswithe}
61 Vir b & 4.2150  & 0.0006  & 166 & 53 & 0.12 & 0.11 & 105 & 54 & 2.12 & 0.23 & 5.1  & 0.5  & 0.050201 & 0.000005 \\
61 Vir c & 38.021  & 0.034   & 177 & 40 & 0.14 & 0.06 & 341 & 38 & 3.62 & 0.23 & 18.2 & 1.1  & 0.2175   & 0.0001    \\
61 Vir d & 123.01  & 0.55    &  56 & 25 & 0.35 & 0.09 & 314 & 20 & 3.25 & 0.39 & 22.9 & 2.6  & 0.476    & 0.001     \\
\enddata
\end{deluxetable}

\end{document}